\documentclass[aip,jpc,twocolumn,reprint]{revtex4-1}
\usepackage{amsmath}
\usepackage{amssymb}
\usepackage{graphicx}
\usepackage{hyperref}
\usepackage{url}
\begin{document}

\title{Superconductivity of monolayer Mo$_2$C: the key role of functional groups}
\author{Jun-Jie Zhang}
\author{Shuai Dong}
\email{sdong@seu.edu.cn}
\affiliation{Department of Physics, Southeast University, Nanjing 211189, China}
\date{\today}

\begin{abstract}
Monolayer Mo$_2$C is a new member of two-dimensional materials. Here the electronic structure and lattice dynamics of monolayer Mo$_2$C are calculated. According to the electron-phonon interaction, it is predicted that monolayer Mo$_2$C could be a quasi-two-dimensional superconductor and the effects of functional-groups are crucially important considering its unsaturated surface. Despite the suppressed superconductivity by chalcogen adsorption, our most interesting prediction is that the electron-phonon interaction of monolayer Mo$_2$C can be greatly enhanced by bromine absorbtion, suggesting that Mo$_2$CBr$_2$ as a good candidate for nanoscale superconductor.
\end{abstract}
\maketitle

\section{Introduction}
Two-dimensional (2D) graphene-like carbides and carbonitrides (MXenes, e.g. Ti$_2$C and Nb$_2$C) have attracted enormous interest for their novel chemical and physical properties, since they were successfully produced by etching the A layers of MAX phase (M is an early transition metal, A is an element from group IIIA or IVA, and X is carbon or nitrogen) \cite{naguib2011two,naguib2012two,naguib2013new}. Due to the unsaturated surface with unpaired electrons, the surfaces of MXenes always easily adsorb various functional groups (e.g. F, O, or/and OH group) during etching, thus the chemical and physical properties are varying with various adsorptions. For this reason, MXenes and their functionalized ones have been widely investigated regarding the magnetism, electronic structures, as well as catalytic properties and energy storage \cite{Lukatskaya1502,zhao2014manipulation,khazaei2013novel,ghidiu2014conductive,sun2016new}.

\begin{figure}
\centering
\includegraphics[width=0.38\textwidth]{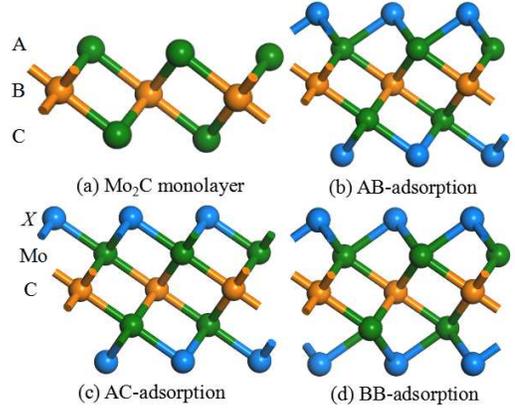}
\caption{Side views of atomic structures of monolayer Mo$_2$C (a) and three atom-adsorbed configurations (b)-(d).}
\label{Fig1}
\end{figure}

Recently, 2D layered Mo$_2$C as a new member of MXenes was formed from Mo$_2$Ga$_2$C thin films. The metallic nature of Mo-Ga bond is weaker than Mo-C bond which has a mixed covalent/metallic/ionic character, thus monolayer Mo$_2$C can be produced by selectively etching Ga layer \cite{meshkian2015synthesis}. Structurally, monolayer Mo$_2$C is constructed by the Mo-C-Mo sandwich, as shown in Fig.~\ref{Fig1}(a), which looks similar to 1T-MoS$_2$. Khazaei \textit{et. al.} studied the thermoelectrics properties of monolayer Mo$_2$C, which was found to be a promising candidate as a high-performance thermoelectric material \cite{khazaei2014two}. In addition, the structural, electrical, thermal and mechanical properties of monolayer Mo$_2$C were also studied \cite{zha2016intrinsic}.

Superconductivity in ultrathin films owns promising future for applications, e.g. superconducting computational devices, thus great efforts have been made to discover 2D superconductors \cite{ge2015superconductivity,zhang2016strain,zhang2016blue}. Considering that monolayer Mo$_2$C is non-magnetic and appears strong metallicity (according to first-principles study) \cite{khazaei2014two}, it brings the opportunity to be an ultrathin superconductor. However, although the superconductivity was discovered in bulk $\alpha$-Mo$_2$C (an allotrope of 2D layered Mo$_2$C), its superconducting transition temperature ($T_{\rm C}$) was depressed from $4$ K (thick film) to near $0$ K when its thickness is less than $3.5$ nm \cite{xu2015large,kolel2007size}. Moreover, analogous to other members of MXenes, the surfaces of monolayer Mo$_2$C are easily covered by functional groups, and Khazaei \textit{et. al.} indicates the full adsorption of functional groups is more stable than the cases of partial adsorption \cite{xie2013hybrid}. For multilayer Mo$_2$C MXenes, they (stacking like MoS$_2$) seem to be impossible due to its unsaturated surfaces. In experiments, the possible stackings are -Mo$_2$C$X_2$-Mo$_2$C$X_2$- and Mo-C-Mo-C-Mo. For the former, the effects of functional groups are unavoidable. For the latter, the stoichiometry has changed. These facts motive us to investigate the superconductivity of pure Mo$_2$C monolayer and those with various functional groups.

In this work, the lattice dynamics and electron-phonon coupling (EPC) of monolayer Mo$_2$C and functionalized ones have been studied via first-principles density functional theory (DFT) and density function perturbation theory (DFPT). Our calculations find that the strong EPC in monolayer Mo$_2$C may lead to superconductivity, while the oxidation would make its superconductivity disappear. Besides, the superconducting $T_{\rm C}$ is also obviously suppressed by absorbtion of sulfur or selenium. In contrast, the EPC of monolayer Mo$_2$C can be greatly enhanced by bromine absorbtion, leading to a predicted $T_{\rm C}$ up to $12$ K.

\section{Model \& methods}
The electronic structure calculations have been performed using the Vienna {\it ab initio} simulation package (VASP) with projector augmented wave method \cite{kresse1996efficient,kresse1999ultrasoft}. Generalized gradient approximation of Perdew-Burke-Ernzerhof (GGA-PBE) are used with a cutoff energy $600$ eV. The spin-orbit coupling (SOC) is also included in electronic structure calculations. The phonon dispersion calculations are carried out using ultrasoft pseudo-potential (including the semicore electrons as valence electrons in case of Mo) as implemented in PWSCF program of the Quantum-ESPRESSO distribution, which are calculated within the framework of DFPT \cite{giannozzi2009quantum}. In the DFPT calculation, GGA-PBE formulation is also used with a cutoff energy $35$ Ry for the expansion of the electronic wave function in the plane waves, whereas cutoff energy for charge density and potential is set to be $350$ Ry. Structure optimization and electronic structure are repeated by using PWSCF and the results are consistent with those obtained using VASP.

The vacuum space of $\sim15$ {\AA} is intercalated into interlamination to eliminate the interaction between layers. A $12\times12$ 2D grid uniform is applied for both the $k$-points of the self-consistent and $q$-points of dynamical matrices calculations.

The EPC calculation is estimated according to the Migdal-Eliashberg theory [$\alpha^{2}F(\omega)$], which is given by \cite{grimvall1981electron}:
\begin{equation}
\alpha^2F(\omega)=\frac{1}{2\pi N(\varepsilon_{F})}\sum_{\bf{q}\nu}\delta (\omega-\omega_{\bf{q}\nu})\frac{\gamma_{\bf{q}\nu}}{\hbar \omega_{\bf{q}\nu}}
\end{equation}
where $N(\varepsilon_{F})$ is the electronic DOS at Fermi level, $\omega_{\bf{q}\nu}$ denotes phonon frequency of the $\nu$th phonon mode with wave vector $\bf{q}$, and the phonon linewidth $\gamma_{\bf{q}\nu}$ is defined by \cite{allen1975transition,allen1972neutron}:
\begin{eqnarray}
\gamma_{\bf{q}\nu}=\frac{2\pi\omega_{\bf{q}\nu}}{\Omega_{BZ}}\sum_{ij}\int &d^3k&\left|g_{\bf{ki},\bf{k+qj}}^{\bf{q}\nu }\right|^2\delta(\varepsilon _{\bf{ki}}-\varepsilon_F)\nonumber\\
&& \times\delta(\varepsilon _{\bf{k+qj}}-\varepsilon_F).
\end{eqnarray}
where $ij$ denote indices of energy bands, $\Omega_{BZ}$ is volume of  Brillouin zone, $\varepsilon _{\bf{ki}}$ and $\varepsilon _{\bf{k+qj}}$ are eigenvalues of Kohn-Sham orbitals at given bands and wave vectors. The $g_{\bf{k},\bf{q}\nu}$ is the EPC matrix element which can be determined self-consistently by the linear response theory, which describes the probability amplitude for scattering of an electron with a transfer of crystal momentum $\bf{q}$, is determined by \cite{allen1975transition,allen1972neutron}:
\begin{equation}
g_{\bf{k},\bf{q}\nu}^{ij}=(\frac{\hbar}{2M\omega_{\bf{q}\nu}})^{1/2}\langle\psi_{i,\bf{k}}|\frac{dV_{SCF}}{d\hat{u}_{\bf{q}\nu}}\cdot\hat{e}_{\bf{q}\nu}|\psi_{i,\bf{k+q}}\rangle
\end{equation}
where $M$ is the atomic mass, $\frac{dV_{SCF}}{d\hat{u}_{\bf{q}\nu}}$ measures the change of self-consistent potential induced by atomic displacement, $\psi_{i,\bf{k}}$ and $\psi_{i,\bf{k+q}}$ are Kohn-Sham orbitals.

The EPC constant $\lambda$ is obtained by summation over the first Brillouin zone or integration of the $\alpha^{2}F(\omega)$ in the $\bf{q}$ space \cite{allen1975transition,allen1972neutron}:
\begin{equation}
\lambda=\sum_{\bf{q}\nu}\lambda_{\bf{q}\nu}=2\int_{0}^{\infty}\frac{\alpha^2F(\omega)}{\omega}d\omega
\label{lambda}
\end{equation}
where EPC constant $\lambda_{\bf{q}\nu}$ for mode $\nu$ at wave vector $\bf{q}$ is defined by the integration \cite{allen1975transition,allen1972neutron}:
\begin{equation}
\lambda_{\bf{q}\nu}=\frac{\gamma_{\bf{q}\nu}}{\pi\hbar N(\varepsilon_{F})\omega_{\bf{q}\nu}^2}.
\label{lambda2}
\end{equation}
To obtain accurate electron-phonon interaction matrices, a dense $36\times36\times1$ grid is adopted for the EPC calculation.

\section{Results \& discussion}
\subsection{Unfunctionalized monolayer Mo$_2$C}
The structure of unfunctionalized monolayer Mo$_2$C is layered hexagonal with a space group of $D_{3d}$, and the stacking of the Mo-C-Mo is in the $ABC$-type along the hexagonal $c$ axis (see Fig.~\ref{Fig1}(a)), which is similar to 1T-MoS$_2$. Our optimized lattice constant ($2.965$ {\AA}) is consistent with the value reported by Khazaei \textit{et. al.} \cite{khazaei2014two}.

\begin{figure}
\centering
\includegraphics[width=0.49\textwidth]{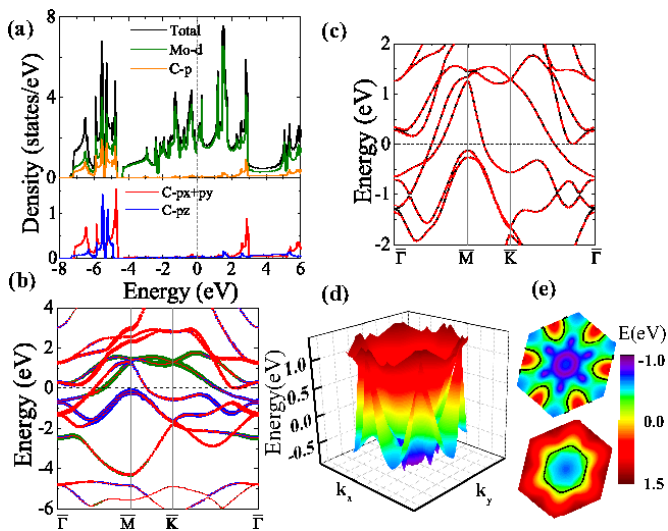}
\caption{Electronic structure of monolayer Mo$_2$C. (a) Density of states (DOS) and Projected DOS. (b) Projected band structure. Red: $d_{z^2}$; Green: $d_{x^2-y^2}$ and $d_{xy}$, Blue: $d_{xz}$ and $d_{yz}$. (c) Band structure with (black) and without SOC (red). (d) Three-dimensional view of two bands cross the Fermi level. (e) Energy couterplots of two bands. Black curves: the Fermi surface.}
\label{Fig2}
\end{figure}

The calculated band structures and density of states (DOS) are shown in Fig.~\ref{Fig2}. As previously reported \cite{khazaei2014two}, monolayer Mo$_2$C indicates strong metallic behavior, and the Fermi surfaces are mainly contributed by Mo's $d$-orbits according to the projected DOS (Fig.~\ref{Fig2}(a)). The maximally localized Wannier functions (MLWFs) can partion Mo's $d$ orbital and C's $p$ orbital, as shown in Fig.~\ref{Fig2}(b). C's $p$ orbital is mainly occupied on energy range from $-8$ eV to $-4$ eV and highly hybridize with Mo's $d$ orbitals. The $p_{x}$ and $p_{y}$ orbitals are degenerate, which are higher in energy than the $p_{z}$ orbital, as shown in Fig.~\ref{Fig2}(a). The Mo's $d$-orbital are split due to $C_{3v}$ crystalline field. Thus the $d_{z^2}$ has the lowest on-site energy, the rests are pairwise degenerate: ($d_{x^2-y^2}$ and $d_{xy}$) and ($d_{xz}$ and $d_{yz}$), the later of which is sightly higher. The SOC can open the degeneracy at $\bar{\Gamma}$ point (see Fig.~\ref{Fig2}(c)), while it has little effect on the Fermi surface. The crossing-Fermi-level bands and corresponding Fermi surfaces are shown in Fig.~\ref{Fig2}(d) and (e). The antibonding $d_{z^2}$ band (the upper one in Fig.~\ref{Fig2}(e)) forms the hole pocket Fermi surfaces around $\bar{M}$ points, while the electron Fermi surface (the lower one in Fig.~\ref{Fig2}(e)) is the circular-shape around the $\bar{\Gamma}$ point which is mainly contributed by $d_{x^2-y^2}$ and $d_{xy}$. Such double Fermi surfaces correspond to carriers with multiple effective masses, charges, and mobilities.

\begin{figure}
\centering
\includegraphics[width=0.45\textwidth]{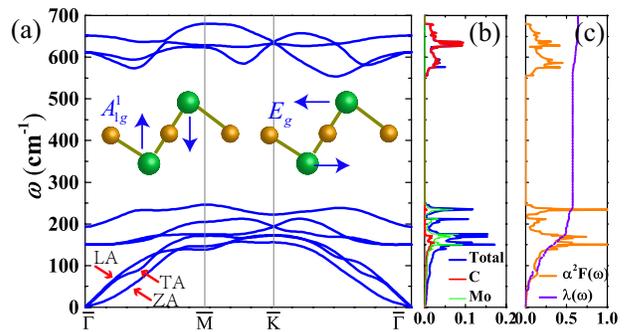}
\caption{Phonon properties for single Mo$_2$C layer. (a) Calculated phonon dispersion. Insert: sketch of Raman modes. (b-c) Phonon DOS, electron-phonon coupling $\lambda$, and Eliashberg spectral function.}
\label{Fig3}
\end{figure}

Then the phonon properties and electron-phonon coupling are calculated. The SOC is not included in these calculations, considering its negligible effects on the Fermi surfaces. For monolayer Mo$_2$C, the Raman modes can be decomposed as $A_{1g}^1 \oplus 2E_{g}^1$ at the zone center (see Fig.~\ref{Fig3}(a)), and the calculated Raman frequencies are $150.3$ cm$^{-1}$ and $193.1$ cm$^{-1}$ for the $E_{g}^1$ and $A_{1g}^1$ modes respectively. Moreover, both Raman modes have strong coupling to electrons ($\lambda_{\bf{q}\nu}$ is $0.11$ for $E_{g}^1$ and $0.21$ for $A_{1g}^1$) according to Eq.~\ref{lambda2}. 

The calculated phonon dispersions along major high symmetry lines and phonon densities of states (PDOS, $F$($\omega$)) for monolayer Mo$_2$C are  shown in Fig.~\ref{Fig3}(a). More dense $k$-meshes ($18\times18\times1$ and $24\times24\times1$) in self-consistent calculations are also tested. The maximum error of obtained phonon frequencies are less than 1\%, implying the convergence of phonon calculation. In addition, the different methodology and pseudo-potential would lead to a few differences regarding the phonon band structures \cite{zha2016intrinsic,sun2016new}. In fact, our phonon structure is very close to that in previous report \cite{zha2016intrinsic}, although tiny differences remain unavoidable.

No imaginary frequency exists in the full phonon spectra, indicating the dynamical stability of monolayer Mo$_2$C. Meanwhile, the phonon behavior exhibits several remarkable characteristics. First, near the zone center, both the LA and TA branches are near linear while the ZA branch (out-of-plane acoustical mode) is quadratic. These characters reflect the nature of 2D sheet. In detail, the ZA phonon in 2D materials like graphene has a quadratic dispersion over a wide range of the 2D Brillouin zone $\omega_{\rm ZA}=a_{\rm ZA}q^2$, where $a_{\rm ZA}$ is a positive constant and $q$ is the 2D phonon wave vector. The similar conclusions are also found in monolayer black phosphorene \cite{Qin2015Anisotropic}. Second, according to the partial PDOS (Fig.~\ref{Fig3}(b)), the vibrational modes of Mo dominate the low-frequency regime while those of C dominate the high-frequency regime, due to their large difference in mass. There is a large gap around $300$ cm$^{-1}$, which partions the optical modes of Mo and C.

To discuss the superconductivity, the obtained $\alpha^{2}F(\omega)$ and $\lambda(\omega)$ are also plotted in Fig.~\ref{Fig3}(c). Their similar shapes indicate that all $F(\omega)$ make contributions to EPC. Due to the factor $1/\omega$ in the definition of $\lambda$ (see Eq.~\ref{lambda}), the contributions from low $\omega$ region is more prominent. Exactly, the calculated $\lambda(\omega=250$ cm$^{-1})$ is $\approx0.56$ which is beyond $90\%$ of the total EPC ($\lambda(\omega=\infty)=0.63$), indicating that the phonon modes in the frequency region below $250$ cm$^{-1}$ have the dominant contribution. In particular, three low-lying optical branches have strong coupling to electrons, which make $40\%$ contribution to EPC. Therefore, it is natural to expect the strong EPC in monolayer Mo$_2$C to induce superconducting state. The $T_{\rm C}$ can be estimated using the Allen-Dynes modified McMillan equation \cite{allen1975transition}:
\begin{equation}
T_{\rm C}=\frac{\omega_{ln}}{1.2}\exp[-\frac{1.04(1+\lambda)}{\lambda-\mu^*(1+0.62\lambda)}],
\label{McMillan}
\end{equation}
where $\mu^*$ is the Coulomb repulsion parameter and $\omega_{ln}$ is the logarithmically averaged frequency. When taking a typical value $\mu^{*}=0.1$, the estimated $T_{\rm C}$ is about $5.9$ K.

\subsection{Functionalized monolayer Mo$_2$C}
In above study, it has been predict that pure monolayer Mo$_2$C maybe a quasi-2D superconductor. However, in real situations, the functional group at surfaces of monolayer Mo$_2$C are unavoidable considering its highly unsaturated surfaces. In fact, previous studies reported that monolayer Mo$_2$C transforms from metal to semiconductor with F- and Cl-adsorption \cite{khazaei2014two}, namely the F- and Cl-functionalized monolayer Mo$_2$C could not be a superconductor. In the following, the changes of superconductivity by adsorbing various functional group will be studied, which may shed light to tuning the superconductivity of monolayer Mo$_2$C in real experiments.

Due to the full adsorption is more stable than the partial case \cite{khazaei2014two}, the $1\times1$ Mo$_2$C unit cell (u.c.) with two functional groups $X$ (one on each surface), i.e. Mo$_2$C$X_2$, is adopted in our calculation. Four functional atoms ($X$=O, S, Se, and Br) are considered. Considering the symmetry, there are three mostly possible site-configuration for adsorption: AB-adsorption (Fig.~\ref{Fig1}(b)), AC-adsorption (Fig.~\ref{Fig1}(c)), BB-adsorption (Fig.~\ref{Fig1}(d)). For the AB-adsorption, one $X$ atom is right above C layer, while another $X$ is right below other side Mo. For the AC-adsorption, each $X$ atom is right above/below other side Mo layer. For the BB-configuration, both two $X$ atoms stand above/below the C site.

\begin{table}
\centering
\caption{The calculated total energy for Mo$_2$C$X_2$ of different configurations for adsorption is in unit of eV/per u.c.. The energy of BB-adsorption is set as the reference. The optimized lattice constant $a$ is in unit of {\AA}. For the most favorable configuration, the corresponding binding energy ($E_b$) (in unit of eV) is also presented.}
\begin{tabular*}{0.45\textwidth}{@{\extracolsep{\fill}}lcccccc}
\hline
\hline
$~$ &AB   &AC    &BB  &$a$  &$a$ (Ref.~\onlinecite{khazaei2014two})  &$E_{b}$ \\
\hline
O &$0.72$     &$1.29$ & $0$  &$2.891$  &$2.886$ &$-8.67$\\
S &$0.48$     &$0.69$ & $0$  &$3.087$ &$3.078$  &$-6.08$\\
Se &$0.39$     &$0.24$ &$0$ &$3.161$  &$~$ &$-4.57$\\
Br &$-0.19$     &$-0.65$ &$0$ &$3.428$  &$3.418$ &$-3.13$\\
\hline
\hline
\end{tabular*}
\label{Table1}
\end{table}

\begin{figure*}
\centering
\includegraphics[width=1.00\textwidth]{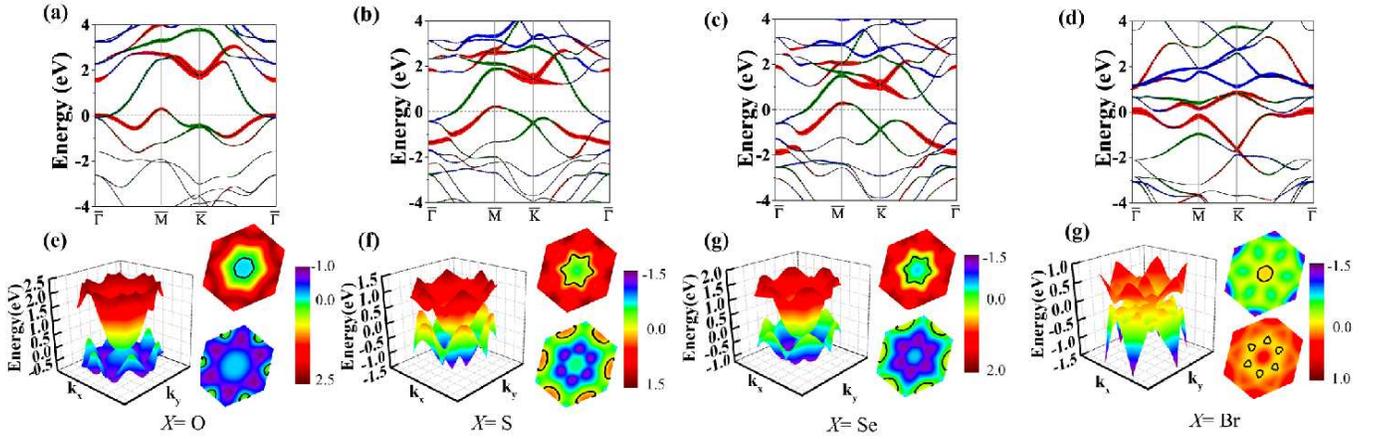}
\caption{Electronic structures of Mo$_2$C$X_2$. (a-d) Band structures. Red: $d_{z^2}$; Green: $d_{x^2-y^2}$ and $d_{xy}$; Blue: $d_{xz}$ and $d_{yz}$. (e-h) Three-dimensional view of three bands around the Fermi level and the corresponding Fermi surface (Black curves).}
\label{Fig4}
\end{figure*}

The crystal structures are fully relaxed upon the absorptions, and the calculated lattice constant as well as total energy are listed in Table~\ref{Table1}. Our calculations are in good agreement with previous reports \cite{khazaei2014two,weng2015large}, and the BB-adsorption is the most favorable case for chalcogen, while Mo$_2$CBr$_2$ favors the AC-adsorption. The corresponding binding energy ($E_{b}=E_{Mo_{2}CX_{2}}-E_{Mo_{2}C}-E_{X_{2}}$) is also calculated and also listed in Table~\ref{Table1}. All binding energies are negative which indicate thermodynamic stability for all structures. And the value of binding energy decreases from Mo$_2$CO$_2$ to Mo$_2$CBr$_2$, implying the adsorption of oxygen should be quite possible, as observed in real experiment \cite{meshkian2015synthesis}. According to  the L\"{o}owdin population analysis, the charge transfer from Mo to $X$ is about $0.48$, $0.39$, $0.34$, and $0.24$ electron for O-, S-, Se-, and Br-adsorption, respectively. Stronger Coulomb attraction between $X$ and Mo ions can lead to larger binding energy.

The electronic structures for Mo$_2$C$X_2$ are calculated. Due to the unchanged symmetry of crystalline field, the splitting of Mo's $d$ orbitals is similar to pristine monolayer Mo$_2$C. Quantitatively, the absorption of $X$ further raises the energy of doubly-degenerate $d_{xz}$ and $d_{yz}$ due to the strong hybridization with the $p$ orbitals of $X$. In details, the on-site energy difference between $d_{xz}$/$d_{yz}$ and $d_{x^2-y^2}$/$d_{xy}$ is about $1.72$, $0.99$, $0.71$, $0.21$ eV for the O-, S-, Se- and Br-adsorption respectively, according to the MLWFs calculation. The lower of $d_{xz}$/$d_{yz}$ orbitals can be also evidenced in the band structures of Mo$_2$C$X_2$, as shown in Fig.~\ref{Fig4}(a-d). All band structures show metallic character after $X$-absorption, different to previously studied F-/Cl-absorption. Similar to pristine monolayer Mo$_2$C, $d_{x^2-y^2}$/$d_{xy}$ and $d_{z^2}$ of Mo still make dominant contributions around the Fermi level, although the contributions from $d_{x^2-y^2}$/$d_{xy}$ are reduced more or less, as shown in Fig.~\ref{Fig4}(a-d). The three-dimensional view of three bands as well as the Fermi surface of $d_{x^2-y^2}$/$d_{xy}$ and $d_{z^2}$ are shown in Fig.~\ref{Fig4}(e-h). For chalcogen, its band structure looks like that of pristine monolayer Mo$_2$C, namely the upper cone-shape band is surrounded by lower flower-shape band, although the degree of surrounding is suppressed. Due to the different adsorption site and valence state of Br, the shapes of Fermi surfaces are significantly changed to sun-like patterns for Mo$_2$CBr$_2$.

\begin{figure}
\centering
\includegraphics[width=0.5\textwidth]{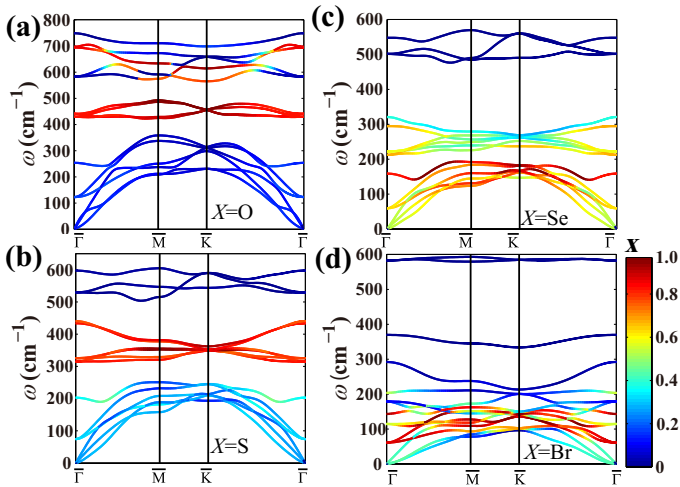}
\caption{Calculated phonon dispersion for Mo$_2$C$X_2$. The contribution from $X$ and monolayer Mo$_2$C is distinguished by color.}
\label{Fig5}
\end{figure}

The calculated phonon dispersion along major high symmetry lines are shown in Fig.~\ref{Fig5}. No imaginary frequency exists in the full phonon spectra, indicating the dynamical stability of the Mo$_2$C$X_2$. Therefore, it is more likely that Mo$_2$C$X_2$ is able to be obtained in real experiment considering its stability in thermodynamics and lattice dynamical. Obviously, the vibration modes of Mo$_2$C are strongly coupled with the surface $X$, as revealed by the phonon dispersion (Fig.~\ref{Fig5}). Different $X$ atom contribute to the phonon spectrum in different frequency range due to the inequivalent mass and bond strength. In particular, for Mo$_2$CO$_2$, the contributions from O mainly locate at the intermediate- and high-frequency regimes, indicating the fact of strong bond of Mo-O. In the case of Mo$_2$CS$_2$, the frequencies contributed by S mainly locate in the intermediate regime, while for Se and Br cases the contributions from $X$ is in the low frequency side.

\begin{figure}
\centering
\includegraphics[width=0.48\textwidth]{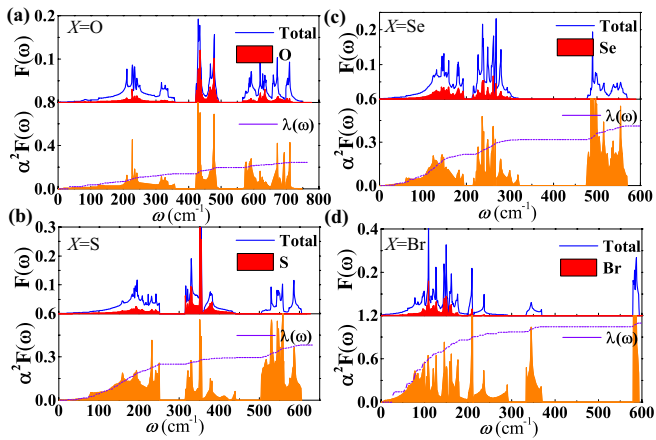}
\caption{Phonon DOS (PDOS, $F(\omega)$), projected PDOS of $X$ atoms, electron-phonon coupling $\lambda(\omega)$, and Eliashberg spectral function of Mo$_2$C$X_2$.}
\label{Fig6}
\end{figure}

Our results for $\alpha^{2}F(\omega)$, $F$($\omega$) and PDOS of Mo$_2$C$X_2$ are shown in Fig.~\ref{Fig6}. As in the monolayer Mo$_2$C case, $\alpha^{2}F(\omega)$ and $F$($\omega$) of Mo$_2$C$X_2$ have similar peaks, indicating all vibration modes contribute to EPC. Comparing to pristine Mo$_2$C, the strength of $\alpha^{2}F(\omega)$ have been suppressed by O-, S- and Se-adsorption. And such suppression obviously exists in low frequency regime which have large contributions to EPC (because of the $\omega^{-1}$ part in Eq.~\ref{lambda}). The average EPC is also calculated using Eq.~\ref{lambda}, as listed in Table~\ref{Table2}. Corresponding $T_{\rm C}$ is estimated from modified McMillan equation (Eq.~\ref{McMillan}) with $\mu^{*}=0.1$ (listed in Table~\ref{Table2}). The results indicate that the superconductivity is greatly suppressed in Mo$_2$CS$_2$ and Mo$_2$CSe$_2$ and almost disappears in Mo$_2$CO$_2$, as a result of reduced electronic DOS's at Fermi level and suppressed EPC's.

\begin{table}
\centering
\caption{The calculated Mo$_2$C$X_2$'s superconductive parameters of $N(\varepsilon_{F})$ (states/eV), $\omega_{ln}$ (K), $\lambda$, and $T_{\rm C}$ (K).}
\begin{tabular*}{0.45\textwidth}{@{\extracolsep{\fill}}lcccc}
\hline
\hline
 $~$                  &Mo$_2$CO$_2$   &Mo$_2$CS$_2$   &Mo$_2$CSe$_2$   &Mo$_2$CBr$_2$  \\
\hline
$N(\varepsilon_{F})$    &$1.3$          &$1.5$          &$1.6$           &$3.3$          \\
$\omega_{ln}$           &$357.4$        &$326.6$        &$283.7$         &$160.7$         \\
$\lambda$               &$0.2$          &$0.4$          &$0.4$           &$1.1$            \\
$T_{\rm C}$             &$<0.1$         &$1.0$          &$1.4$           &$12.8$            \\
\hline
\hline
\end{tabular*}
\label{Table2}
\end{table}

As shown in Fig.~\ref{Fig6}(d), strength of $\alpha^{2}F(\omega)$ has been improved in Mo$_2$CBr$_2$, and the corresponding average EPC is about $1.09$. The obtained $\lambda(\omega=300$ cm$^{-1})\approx0.97$ is approximately beyond $88\%$ of the total EPC ($\lambda(\omega=\infty)=1.09$), and estimated $T_{\rm C}$ is up to $12.8$ K. To be exact, EPC of vibration modes ($E_{g}^1$ and $A_{1g}^1$) at $\bar{\Gamma}$ point which have Raman activity have been improved for $30\%$ comparing to those of pristine monolayer Mo$_2$C. In addition, the large values of EPC for acoustic modes at $\bar{M}$ contribute  substantially to the average EPC. Thus, the superconductive $T_{\rm C}$ is pushed up in Mo$_2$CBr$_2$.

In addition, we had tried the full -OH cover up Mo$_2$C surfaces. However, even after the atomic relaxation, an imaginary frequency of phonon appears at the $\bar{M}$ point, implying unstable structure. Therefore, structural phase transition would appear which makes the problem more complicated. Thus, the decoration of OH group is beyond the current work and deserves individual studies in future.

\section{Conclusion}
We have analyzed the electronic properties, the lattice dynamical properties, and superconductivity of monolayer Mo$_2$C and its functionalized ones. Our calculations have confirmed the strong EPC in monolayer Mo$_2$C, which may lead to superconductivity below $5.9$ K. Even though, since the absorption of functional groups is unavoidable in real experiment, its superconductivity can be modified. Our calculation have found that for chalcogen functionalized monolayer Mo$_2$C the superconductivity would be seriously suppressed (or even totally disappear). The most interesting prediction is that electron-phonon coupling can be greatly enhanced in monolayer Mo$_2$C by bromine absorbtion, and thus its corresponding superconductive $T_{\rm C}$ may be pushed up to $12.8$ K, suggesting that Mo$_2$CBr$_2$ may be a good candidate as nanoscale superconductor.

\begin{acknowledgments}
Work was supported by the National Natural Science Foundation of China (Grant No. 11674055), the Fundamental Research Funds for the Central Universities, Jiangsu Innovation Projects for Graduate Student (Grant No. KYLX16\underline{ }0116).
\end{acknowledgments}
\bibliographystyle{apsrev4-1}
\bibliography{ref}
\end{document}